\documentclass[11pt,a4paper]{article}
\usepackage{jheppub}

\usepackage[dvipsnames]{xcolor}
\usepackage{bbm}

\def\eps{\epsilon}

\def\be{\begin{equation}}
\def\ee{\end{equation}}

\newcommand{\oeps}[1]{O(\eps^{#1})}

\author{John Golden}
\author{and Daniel R. Mayerson}
\title{Mellin Bootstrap for Scalars in Generic Dimension}

\affiliation{Department of Physics and Leinweber Center for Theoretical Physics,\\
University of Michigan, 450 Church Street, Ann Arbor, MI 48109-1020, USA}

\emailAdd{jkgolden@umich.edu}
\emailAdd{drmayer@umich.edu}

\abstract{
We use the recently developed framework of the Mellin bootstrap to study perturbatively free scalar CFTs in arbitrary dimensions. This approach uses the crossing-symmetric Mellin space formulation of correlation functions to generate algebraic bootstrap equations by demanding that only physical operators contribute to the OPE. We find that there are no perturbatively interacting CFTs with only fundamental scalars in $d>6$ dimensions (to at least second order in the perturbation). Our results can be seen as a modest step towards understanding the space of interacting CFTs in $d>6$ and are consistent with the intuition that no such CFTs exist.
}
\preprint{LCTP-17-05}

\begin{document}

\maketitle

\section{Introduction and Summary}

The program of the conformal bootstrap was introduced in the 1970s~\cite{Polyakov:1974gs,Ferrara:1973yt} and has recently regained attention~\cite{Rattazzi:2008pe,Rychkov:2016iqz,Simmons-Duffin:2016gjk,Poland:2016chs}. This program is largely based on the fact that the consequences of conformal symmetry are sufficiently stringent to significantly constrain the space of possible CFTs. Two particularly important consequences are the operator product expansion (OPE) and crossing symmetry. Considerable insights can be gained by studying the constraints of crossing symmetry on theories which, by construction, satisfy the OPE (e.g. \cite{ElShowk:2012ht}). This is done by constructing an ansatz in the form of an OPE of a four-point correlation with generic coefficients, and then imposing crossing symmetry and exploring the solution space. The recent \emph{Mellin bootstrap}, pionereed by Gopakumar, Kaviraj, Sen, Sinha in \cite{Gopakumar:2016wkt,Gopakumar:2016cpb},\footnote{Some earlier works important for the development of the Mellin bootstrap are \cite{Mack:2009mi,Penedones:2010ue,Costa:2012cb}.} approaches the problem from the complementary perspective of assuming crossing symmetry and then imposing the correct OPE structure. In this case, one constructs an ansatz that explicitly satisfies crossing symmetry but then the correct operator spectrum has to be imposed as a consistency condition. Demanding that the ``exchanged'' operators in the four-point correlation function are physical operators produces an infinite set of algebraic constraints. 

For the $d=4-\eps$ and $d=6-\eps$ Wilson-Fisher fixed points studied in \cite{Gopakumar:2016wkt,Gopakumar:2016cpb,Dey:2017oim}\footnote{A similar $\eps$-expansion analysis for the $O(N)$ model was made using the Mellin bootstrap in \cite{Dey:2016mcs}; the large-spin structure of the Mellin bootstrap method was analyzed in \cite{Dey:2017fab} without reference to a particular $\eps$-expansion.}, it was shown that, to a certain order in $\eps$, this infinite set of algebraic equations in the Mellin bootstrap truncates to an analytically solvable finite set. It is interesting to consider whether this simplification of the Mellin bootstrap is a feature of perturbatively free CFTs in general. In this note, we begin by investigating theories with a single fundamental scalar $\phi$ that are perturbatively close to a free point in a general dimension $d$ (which can be a function of the perturbation parameter $\eps$ as in the Wilson-Fisher fixed point but does not necessarily have to be). We show that the Mellin bootstrap again reduces to a finite set of solvable analytic equations in this more general case. Furthermore, the solutions to these equations constrains the dimensions of $\phi$ and the lowest excited scalar operator ``$\phi^2$", as well as the OPE coefficients $C_{\phi\phi T}$, $C_{\phi\phi\phi}$ and $C_{\phi\phi\phi^2}$. In particular, we find that the scalar theory is \emph{forced} to be the free scalar theory in arbitrary dimension $d>6$, to the perturbative order we are able to calculate. This provides evidence that there are no interacting CFTs with a single fundamental scalar in $d>6$ dimensions. Our results can also easily be generalized to the case of an arbitrary number of fundamental scalars, again providing evidence that there are no interacting CFTs with \emph{any number of} fundamental scalars in $d>6$ dimensions.

Our results are perhaps not surprising: for perturbatively free scalar theories, Lagrangian methods give an easy argument that no non-trivial marginal operator can be constructed in $d>6$ for scalar theories - therefore no interacting scalar theory would be expected to exist. Furthermore, recall that the superconformal algebra does not close in $d>6$, so there are no superconformal field theories in in this regime~\cite{Nahm:1977tg}. However, the Mellin bootstrap approach allows us to approach the question of existence of CFTs from a fundamentally different viewpoint not involving Lagrangians at all. This approach also begins to clarify the criteria under which the Mellin bootstrap reduces to a finite set of equations. It would be interesting to expand our results to include fermions, other (fundamental) fields, and to connect these techniques with the analysis of generalized Wilson-Fisher fixed points in \cite{Gliozzi:2016ysv, Gliozzi:2017hni}.

This paper is organized as follows: section \ref{sec:setup} briefly introduces the Mellin bootstrap and our CFT setup, including our assumptions and the resulting simplifications that occur that reduce the bootstrap equations to a finite number of algebraic equations. Section \ref{sec:bootstrap} discusses solving these bootstrap equations to show that the theory we obtain is forced to be free. The necessary definitions of functions etc. can be found in appendix \ref{sec:app:def} while the simplifications are discussed in more detail in appendix \ref{sec:app:simpl}.

\section{Setup, Assumptions, and Simplifications} \label{sec:setup}
The objects of study for this note are unitary, perturbative CFTs containing one fundamental scalar $\phi$ as the lowest-dimension operators. (We will discuss in section \ref{sec:multiplescalars} how to extend our results to an arbitrary number of fundamental scalars.)

The CFT we consider depends on a perturbative expansion parameter $\eps$ and lives in $d_\eps$ dimensions (the subscript $\eps$ is to emphasize the fact that $d$ can be an analytic function of $\eps$, as is the case in the Wilson-Fisher CFT). The $\eps=0$ limit corresponds to the free scalar CFT in $d_0$ dimensions. The dimension of $\phi$ is parameterized as
\begin{equation} \label{eq:defdelphi} 
	\Delta_{\phi} = \frac{d_\eps-2}{2} + \delta_{\phi}^{(1)}\eps + \delta_{\phi}^{(2)}\eps^2 + \oeps{3}.
\end{equation} 
When $\delta_{\phi}^{(i)}=0$ for all $i$, $\phi$ saturates the unitary bound and is a free scalar for any $\eps$~\cite{Mack:1975je}. We have a similar expansion for the dimension of the lowest excited operator, which we denote as ``$\phi^2$" following standard convention:
\begin{equation} \label{eq:defdelphi2} 
	\Delta_{\phi^2} \equiv \Delta_0 = d_\eps - 2 + \delta_0^{(1)}\eps + \delta_0^{(2)}\eps^2 + \oeps{3}.
\end{equation}
We also expand the three-point OPE coefficients $C_{\phi\phi\phi}$, $C_{\phi\phi\phi^2}$, and $C_{\phi\phi T}$ ($T$ being the stress-tensor):
\begin{align} 
	\label{eq:defCphi} C_{\phi\phi\phi} &\equiv C_\phi = C_\phi^{(0)} + C_\phi^{(1)}\eps + C_\phi^{(2)}\eps^2 + \oeps{3},\\
	\label{eq:defCphi2}C_{\phi\phi\phi^2} &\equiv C_0 = C_0^{(0)} + C_0^{(1)}\eps + C_0^{(2)}\eps^2 + \oeps{3},\\
	\label{eq:defCT} C_{\phi\phi T} &\equiv C_2 = C_2^{(0)} + C_2^{(1)}\eps + C_2^{(2)}\eps^2 + \oeps{3}.
\end{align}
We will also consider the lowest dimension (primary) operator $J_l$ of spin $l$ (for $l\neq 0,2$), which is given by the schematic form\footnote{We have in mind the symmetric, traceless current operator of spin $l$ that is a conserved current in the free theory. The exact form of these operators can be found in e.g. \cite{Mikhailov:2002bp}.} $\phi \partial_{i_1}\cdots \partial_{i_l}\phi$ and has dimension $\Delta_{J_l}$ and OPE coefficient $C_{\phi\phi J_l}$ given by:
\begin{align}
 \label{eq:defdelJ} \Delta_{J_l} &= d_\eps-2 + l + \delta_l^{(1)}\eps + \oeps{2},\\
 \label{eq:defCJ} C_{\phi\phi J_l} &\equiv C_l = C_l^{(0)} + C_l^{(1)}\eps + \oeps{2}.
 \end{align}

We make the following assumptions about the theory:
\begin{itemize}
	\item \textbf{A1}: There is a conserved stress tensor with $\Delta=d_\eps$ and spin $l=2$ (i.e. the operator $J_{l=2}$ has $\delta^{(i)}_{l=2}=0$ for all $i$).
	\item \textbf{A2}: There is only one fundamental scalar, $\phi$, of dimension $\Delta_\phi$ given in (\ref{eq:defdelphi}).)
\end{itemize}
Note that we do not assume any other symmetry of the theory, such as an overall $\mathbb{Z}_2$, as is often done. We will instead keep our analysis generic and then note where a $\mathbb{Z}_2$ symmetry (i.e. $C_{\phi\phi\phi}=0$) allows us to make even stronger claims. We do note that we have assumed that all relevant quantities in (\ref{eq:defdelphi2})-(\ref{eq:defCJ}) allow for an analytic expansion in $\eps$; this is an assumption we make on the nature of the perturbative expansion.\footnote{We thank M. Paulos for stressing this point to us.} We do not need to assume that $\eps$ is positive; in our analysis we will see that the only possible solution for $d_0>6$ is the free theory
which saturates the unitarity bound in $d_\eps$ dimensions, regardless of the sign of $\eps$. (By contrast, note that $\eps>0$ is necessary in the $d= 6 - \eps$ analysis of \cite{Gopakumar:2016cpb}; there, when $\eps<0$, the resulting $\Delta_\phi$ can be seen to violate the unitarity bound already at $\oeps{1}$.)

Our goal is to determine the values of the expansion coefficients in eqs.~(\ref{eq:defdelphi})-(\ref{eq:defCT}) given the assumptions \textbf{A1} and \textbf{A2}. We will do so using the Mellin bootstrap, which we review now. 

\subsection{The Mellin Bootstrap}

Our analysis is an extension of the Mellin bootstrap techniques introduced in~\cite{Gopakumar:2016wkt} and described in detail in~\cite{Gopakumar:2016cpb}. We are interested in the four-point function (with identical scalars) $\langle\phi\phi\phi\phi\rangle$, which in position space is determined completely by the function $\mathcal{A}(u,v)$ through the relation:
\be \langle \phi(x_1)\phi(x_2)\phi(x_3)\phi(x_4)\rangle = x_{12}^{-2\Delta_\phi} x_{34}^{-2\Delta_\phi} \mathcal{A}(u,v) ,\ee
where $u$ and $v$ are the standard conformal cross ratios:
\be u = \frac{x_{12}^2 x_{34}^2}{x_{13}^2x_{24}^2}, \qquad v = \frac{x_{14}^2 x_{23}^2}{x_{13}^2x_{24}^2}.\ee
As described in the Introduction, the Mellin bootstrap involves constructing $\langle\phi\phi\phi\phi\rangle$ with crossing symmetry explicitly satisfied, but without explicit agreement with the OPE. In other words, it is not guaranteed that the terms that appear in the expansion are primaries and their descendants. Polyakov introduced this crossing symmetric construction in~\cite{Polyakov:1974gs} and noted that in this expansion terms proportional to $u^{\Delta_\phi}$ and $u^{\Delta_\phi}\log u$ necessarily appear. These terms correspond to an operator of dimension $2\Delta_\phi$ appearing in the spectrum. Since there is not generically an operator of this dimension in an interacting CFT, these terms must cancel when summed over all channels. Requiring that this is the case imposes non-trivial constraints on the CFT. However, actually calculating the coefficients of the spurious $u^{\Delta_\phi}$ and and $u^{\Delta_\phi}\log u$ terms is very difficult in position space, so this approach did not recieve much attention until Gopakumar, Kaviraj, Sen, and Sinha realized~\cite{Gopakumar:2016wkt} that the spurious terms in position space become spurious poles, with calculable coefficients, in Mellin space. We now briefly describe this calculational scheme. Appendix \ref{sec:app:def} contains explicit definitions for all of the functions that follow.

The (reduced) Mellin amplitude ${\cal M}(s,t)$ of the four-point function $\mathcal{A}(u,v)$ is defined by
\begin{equation}\label{eq:defmellintransform}
	\mathcal{A}(u,v) = \int_{-i\infty}^{i\infty}\frac{ds}{2\pi i} \, \frac{dt}{2\pi i} \, u^{s}v^t \Gamma^2(-t)\Gamma^2(s+t)\Gamma^2(\Delta_{\phi}-s){\cal M}(s,t),
\end{equation}
where the form of the measure has been chosen to highlight the pole at $s=\Delta_\phi$. In particular, the expansion of $\mathcal{M}(s,t)$ around $s=\Delta_\phi$ has a constant and linear term in $(s-\Delta_\phi)$, so the overal integral has a single and double pole at this point. The residues at these poles give rise to precisely the $u^{\Delta_\phi}$ and $u^{\Delta_\phi}\log u$ terms that indicate the unphysical operator described in the previous paragraph. Therefore, requiring that the coefficients of the $u^{\Delta_\phi}$ and $u^{\Delta_\phi}\log u$ terms be zero then corresponds to ensuring that the constant and linear terms vanish:
\begin{equation}\label{eq:m-coeffs-vanish}
	\mathcal{M}(s=\Delta_\phi,t) = 0 \qquad \text{and} \qquad \left.\frac{\partial}{\partial s}\mathcal{M}(s,t)\right|_{s=\Delta_\phi} = 0.
\end{equation}
This is the most general representation of \textbf{`the Mellin bootstrap constraints'}. While we will not make use of it here, it is interesting to note that there are in fact an infinite class of these constraints, as $\Gamma^2(\Delta_{\phi}-s)$ has spurious poles at $s=\Delta_\phi + n, \forall ~n\in \mathbb{Z}^+$.

Now we must cast $\mathcal{M}(s,t)$ in a form where we can impose (\ref{eq:m-coeffs-vanish}) for all values of $t$. First off, we can decompose $\mathcal{M}(s,t)$ into what can be thought of as exchange (Witten) diagrams $M_{\Delta,l}^{(s/t/u)}(s,t)$ where an operator $\mathcal{O}_{\Delta, l}$ is exchanged between pairs of $\phi$'s; these exchanges can happen in each of the $s,t,u$-channels. This gives us the explicitly crossing symmetric representation
\begin{align}
	\mathcal{M}(s,t) &= \sum_{\Delta, l}c_{\Delta,l}\left(M_{\Delta,l}^{(s)}(s,t) +  M_{\Delta,l}^{(t)}(s,t)+ M_{\Delta,l}^{(u)}(s,t)\right).
\end{align}
Note the introduction of the factor $c_{\Delta,l}$, which is proportional to the square of the OPE coefficient between $\phi\phi$ and $\mathcal{O}_{\Delta, l}$ (from the vertices of the exchange diagram).

Fortunately there is a convenient basis of orthogonal functions, the continuous Hahn polynomials $\{Q^{\Delta}_l(t)\}$, which allow us to partially separate the $s-$ and $t-$dependence in $M^{(s/t/u)}(s,t)$. In this decomposition, only a single term contributes in the $s$-channel:
\begin{equation} 
	\label{eq:Ms} M_{\Delta,l}^{(s)}(s,t) = q_{\Delta,l}^{(s)}(s)Q_l^{2\Delta_\phi+l}(t),
\end{equation}
where
\begin{equation} 
	q_{\Delta,l}^{(s)}(s) = q_{\Delta,l}^{(2,s)} + (s-\Delta_\phi)q_{\Delta,l}^{(1,s)} + \ldots (\text{as }s\to\Delta_\phi).
\end{equation}
$q_{\Delta,l}^{(2,s)}$ and $q_{\Delta,l}^{(1,s)}$ are labeled this way as they correspond to the double and single poles in (\ref{eq:defmellintransform}), respectively. 

The contributions from the $t-$ and $u-$channels turn out to be identical, so we need only include the $t$-channel twice. The infinite number of contributions in the $t-$channel which must cancel against the $s$-channel contribution are:
\begin{equation}
	\label{eq:Mt} M_{\Delta,l}^{(t)}(s,t) = \sum_{l'} q_{\Delta,l|l'}^{(t)}(s) Q_{l'}^{2\Delta_\phi+l'}(t),
\end{equation}
where also:
\begin{equation} 
 	q_{\Delta,l|l'}^{(t)}(s) = q_{\Delta,l|l'}^{(2,t)} + (s-\Delta_\phi)q_{\Delta,l|l'}^{(1,t)} + \ldots (\text{as }s\to\Delta_\phi).
\end{equation}
Note that there is a polynomial ambiguity that we have omitted from (\ref{eq:Ms}) and (\ref{eq:Mt}), which stems from an ambiguity in the Mellin space Witten diagrams related to contact terms \cite{Costa:2012cb,Gopakumar:2016cpb,Dey:2017fab,Dey:2017oim}. This ambiguity was shown to lead to different results for certain quantities in the Mellin bootstrap approach compared to the regular bootstrap approach at high orders in $\eps$ \cite{Dey:2017fab}. While the presence of this ambiguity is thus certainly important, it is also expected that it will only start contributing substantially to the relevant Mellin bootstrap results at higher orders in $\eps$ than we will be considering (specifically, $\oeps{4}$) \cite{Dey:2017fab}.

 Summing up the contributions from the different channels then gives us a more concrete formulation of the \textbf{bootstrap equations}
\begin{equation}\label{eq:bootstrap}
	\sum_{\Delta}\left( c_{\Delta,l} q_{\Delta,l}^{(a,s)} + 2 \sum_{l'}c_{\Delta,l'} q^{(a,t)}_{\Delta,l|l'}\right) = 0,
\end{equation}
where $a=1$ corresponds to the constraint $\left.\frac{\partial}{\partial s}\mathcal{M}(s,t)\right|_{s=\Delta_\phi} = 0$ and $a=2$ corresponds to the constraint $\mathcal{M}(s=\Delta_\phi,t) = 0$.

\subsection{Simplifications}\label{sec:main:simpl}

The bootstrap (\ref{eq:bootstrap}) represents an infinite number of equations (one for each $l$), and each equation involves an infinite number of terms (from the sums over $\Delta,l'$). A number of non-trivial simplifications occur for every $l$, resulting in only a finite number of terms contributing up to $\oeps{2}$. These terms involve the unknown parameters in eqs.~(\ref{eq:defdelphi})-(\ref{eq:defCT}) and form a solvable finite system of equations. These simplifications are:
\begin{itemize}
	\item \textbf{S0}: The identity operator only contributes to the simple pole constraint, so we can separate out that contribution and write the bootstrap equations as
	\begin{align}
		\label{eq:bootstrap2} \sum_{\Delta\neq0}\left( c_{\Delta,l} q_{\Delta,l}^{(2,s)} + 2 \sum_{l'}c_{\Delta,l'} q^{(2,t)}_{\Delta,l|l'}\right) &= 0,\\
		\label{eq:bootstrap1} 2q_{\Delta=0,l|0}^{(1,t)} + \sum_{\Delta\neq0}\left( c_{\Delta,l} q_{\Delta,l}^{(1,s)} + 2 \sum_{l'}c_{\Delta,l'} q^{(1,t)}_{\Delta,l|l'}\right) &= 0.
	\end{align}
	\item \textbf{S1}: In the $s$-channel, only the lowest dimension operator of spin $l$ contributes to $\sum_{\Delta}c_{\Delta,l}q_{\Delta,l}^{(2,s)}$ up to $O(\eps^2)$ (for $l=0$, we will consider the \emph{two} lowest operators explicitly). For $l=2$ the infinite sum reduces to the single term
  	\begin{equation}
  		\sum_{\Delta\neq0}c_{\Delta,l=2}q_{\Delta,l=2}^{(a,s)} = c_{\Delta=d_\eps,l=2}q_{\Delta=d_\eps,l=2}^{(a,s)}+\oeps{2},
  	\end{equation}
  	as the lowest dimension operator of spin $l=2$ is the stress tensor with $\Delta=d_\eps$. For $l=0$ the infinite sum reduces to
  	\be
		\sum_{\Delta\neq0}c_{\Delta,l=0}q_{\Delta,l=0}^{(a,s)} = c_{\Delta=\Delta_\phi,l=0}q_{\Delta=\Delta_\phi,l=0}^{(a,s)} +  c_{\Delta=\Delta_0,l=0}q_{\Delta=\Delta_0,l=0}^{(a,s)}+\oeps{2}
	\ee
	as the two lowest dimension scalars that can contribute are $\phi$ itself and $\phi^2$. For other $l$, the lowest dimension operator is the operator $J_l$, so the infinite sum reduces to:
	\be
	\sum_{\Delta\neq0}c_{\Delta,l}q_{\Delta,l}^{(a,s)} = c_{\Delta,l}q_{\Delta=\Delta_{J_l},l}^{(a,s)}+\oeps{2}.
	\ee
	\item \textbf{S2}: In the $t$-channel, for $l'>0$, we have $c_{\Delta,l'}q^{(a,t)}_{\Delta, l|l'}= \oeps{2}$; i.e. to this order in $\eps$, only scalars contribute to the sum $\sum_{l'} c_{\Delta,l'}q^{(a,t)}_{\Delta, l|l'}$. This infinite sum over $l'$ thus reduces to:
  	\begin{equation} 
  		\sum_{l'} c_{\Delta,l'}q^{(a,t)}_{\Delta, l|l'} = c_{\Delta,l'=0} q^{(a,t)}_{\Delta,l|l'=0} + \oeps{2}.
  	\end{equation}

	\item \textbf{S3}: Furthermore, in the $t$-channel, for ``heavier'' scalars of $\Delta > \Delta_0$, we have $c_{\Delta,0}q^{(a,t)}_{\Delta, l|0}= \oeps{2}$. This reduces the infinite sum over $\Delta$ in the $t$-channel to:
	\be
	\sum_{\Delta} c_{\Delta,l'=0} q^{(a,t)}_{\Delta,l|l'=0} = c_{\Delta=\Delta_\phi,l'=0} q^{(a,t)}_{\Delta=\Delta_\phi,l|l'=0} +  c_{\Delta=\Delta_0,l'=0} q^{(a,t)}_{\Delta=\Delta_0,l|l'=0}+\oeps{2}.
	\ee
\end{itemize}
Derivations of these simplifications are in appendix \ref{sec:app:simpl}. With these simplifications, the bootstrap equations for generic $l$ become
\begin{align}
	\label{eq:ll2} c_{\Delta_{J_l},l} q_{\Delta_{J_l},l}^{(2,s)} + 2 c_{\Delta_\phi,0} q^{(2,t)}_{\Delta_\phi,l|0} +2 c_{\Delta_0,0} q^{(2,t)}_{\Delta_0,l|0} &=\oeps{2}, & (l\neq0,2),\\
		\label{eq:ll1} 2q^{(1,t)}_{0,l|0} +c_{\Delta_{J_l},l} q_{\Delta_{J_l},l}^{(1,s)} + 2 c_{\Delta_\phi,0} q^{(1,t)}_{\Delta_\phi,l|0} +2 c_{\Delta_0,0} q^{(1,t)}_{\Delta_0,l|0} &=\oeps{2},& (l\neq0,2),
\end{align}
and in the special cases $l=0$ and $l=2$ we have
\begin{align}
	\label{eq:l22} c_{d_\eps,2} q_{d_\eps,2}^{(2,s)} + 2 c_{\Delta_\phi,0} q^{(2,t)}_{\Delta_\phi,2|0} +2 c_{\Delta_0,0} q^{(2,t)}_{\Delta_0,2|0} &=\oeps{2},\\
	\label{eq:l21} c_{d_\eps,2} q_{d_\eps,2}^{(1,s)} + 2q_{0,2|0}^{(1,t)} +  2 c_{\Delta_\phi,0} q^{(1,t)}_{\Delta_\phi,2|0} + 2 c_{\Delta_0,0} q^{(1,t)}_{\Delta_0,2|0} &=\oeps{2},\\
	\label{eq:l02} c_{\Delta_\phi,0}\left(q_{\Delta_\phi,0}^{(2,s)}+2 q^{(2,t)}_{\Delta_\phi,0|0}\right)+c_{\Delta_0,0}\left(q_{\Delta_0,0}^{(2,s)} + 2 q_{\Delta_0,0|0}^{(2,t)}\right) &= \oeps{2},\\
	\label{eq:l01} 2q^{(1,t)}_{0,0|0} + c_{\Delta_\phi,0}\left(q_{\Delta_\phi,0}^{(1,s)}+2  q^{(1,t)}_{\Delta_\phi,0|0}\right)+c_{\Delta_0,0}\left(q_{\Delta_0,0}^{(1,s)} + 2 q_{\Delta_0,0|0}^{(1,t)}\right) &= \oeps{2}.
\end{align}

All of the equations now only involve a finite number of terms. Demanding that the left-hand side of these equations vanish up to the given order in $\eps$ will give us non-trivial constraints on the coefficients appearing in (\ref{eq:defdelphi})-(\ref{eq:defCJ}).

\section{Bootstrap}\label{sec:bootstrap}

In this section we describe the bootstrap in considerable detail in order to highlight the differences between the generic $d_\eps$ case and the special $d_\eps = 4-\eps$ and $d_\eps = 6-\eps$ cases\footnote{$d_0=2$ is also obviously a special case, but for the sake of brevity we will not comment on those subtleties here.} studied in \cite{Gopakumar:2016cpb}. Before we do so, let us describe our result in physical terms, as the preceeding notation is a bit cumbersome.

Consider, as an example, the schematic form of (\ref{eq:l21}):
\begin{equation}
	T_\text{($s$-channel)} + \mathbbm{1}_\text{$t$-channel} + \phi_\text{($t$-channel)}+\phi^2_\text{($t$-channel)}= \oeps{2} 
\end{equation}
where by ``$\mathcal{O}_\text{(channel)}$'' we refer to an overall contribution of the form $c_{\Delta_\mathcal{O},l}q_{\Delta_\mathcal{O},l}^{\text{channel}}$. The other bootstrap equations take a similar form, with the $s$-channel contribution coming from either $\phi$ in the $l=0$ case, $T$ for $l=2$, and generic $J_l$ for $l>2$, and the $t$-channel identity operator only arises in the simple-pole constraints. For $d_0\le6$, all of these terms can make finite, non-zero contributions through $\oeps{1}$ (though note that they do not all contribute in all cases -- for example $\phi_\text{($t$-channel)} = 0$ in the $\mathbb{Z}_2$-symmetric case as $C_{\phi\phi\phi}=0$). The essential story for the $d_0\le6$ regime is that the complicated interplay between non-trivial $s$- and $t$-channel contributions produces a set of linear equations to get cancellations up to $\oeps{2}$. This set of linear equations can then be used to determine operator dimensions, e.g. for the Wilson-Fisher fixed point in $d_\eps = 4-\eps$ and the $\phi^3$ theory in $d_\eps = 6-\eps$. 

We will show that for $d_0>6$, the $\phi$ and $\phi^2$ $t$-channel contributions necessarily begin $\oeps{2}$, so they drop out of the bootstrap equations and all we are left with is the $t$-channel identity to cancel the $s$-channel operator. Without these non-trivial $t$-channel contributions, the only possible solution to the resulting set of equations is the free theory. 

\subsection{\texorpdfstring{$t$}{t}-Channel Contributions from \texorpdfstring{$\phi^2$}{phi-squared}}\label{sec:bootstrapphisq}
While we have emphasized $d_0>6$ so far, the $\phi^2$ $t$-channel contribution in fact drops out already at $d_0>4$. We can see from a simple series expansion that the normalization coefficient $c_{\Delta_0,0}$ already exhibits a special value at $d_0=4$:
\begin{equation}
c_{\Delta_0,0} \propto
	\begin{cases}
		\oeps{4} & d_0 = 4,\\
		\Gamma\left(d_0/2-2\right)^{-2}\eps^{2} & d_0 > 4.
	\end{cases}
\end{equation}
A less trivial calculation (see appendices \ref{sec:app:def} and \ref{sec:app:simpl} for details) reveals that
\begin{equation}
q_{\Delta_0,l|0}^{(a,t)}\propto
	\begin{cases}
		\oeps{-3} & d_0 = 4,\\
		\Gamma\left(d_0/2-2\right)^{3}\eps^{0} & d_0 > 4.
	\end{cases}
\end{equation}
Putting these together gives
\begin{equation}\label{eq:divergentterm}
c_{\Delta_0,0}q_{\Delta_0,l|0}^{(a,t)} \propto
	\begin{cases}
		\oeps{1} & d_0 = 4,\\
		\Gamma(d_0/2-2)\eps^2 & d_0 > 4.
	\end{cases}
\end{equation}

The fact that these poles behave differently in $d_0>4$ is not particularly surprising. We know from~\cite{Gopakumar:2016cpb} that at least in $d_0=6$, the $\phi^2$ operator indeed does only contribute at higher order (otherwise it would compete with the $C_{\phi\phi\phi}$ exchange in the non-$\mathbb{Z}_2$ invariant theory); however, what we find here is that the $\phi^2$ operator cannot contribute to the relevant orders for \emph{any} $d_0>4$ (in particular $d_0=4n$ for $n>1$ is not special in any way).

\subsection{\texorpdfstring{$t$}{t}-Channel Contributions from \texorpdfstring{$\phi$}{phi}}\label{sec:bootstrapphi}
Now, we turn to the contributions from $\phi$ in the bootstrap equations. Once again, we first note that a simple series expansion shows that the normalization coefficient $c_{\Delta_\phi,0}$ exhibits a special value at $d_0=6$:
\begin{equation}\label{eq:divergenttermphi1}
c_{\Delta_\phi,0} \propto
	\begin{cases}
		\oeps{1} & d_0 = 6,\\
		\Gamma\left(\frac{d_0-6}{4}\right)^{-2}\eps^{-1} & d_0 > 6.
	\end{cases}
\end{equation}
However, for $d_0>6$, $q^{(a,t)}_{\Delta_0,l|0}$ has an infinite sum of contributions from poles of the integrand of that go as $\oeps{0}$. In particular, this infinite sum is not convergent. This means that in $d_0>6$, we have (schematically):
\be c_{\Delta_\phi,0}q^{(a,t)}_{\Delta_0,l|0} = C_\phi^{(0)} \eps^{-1}(\infty),\ee
where we have made explicit the factor of $C_{\phi}^{(0)}$ that is included in the normalization $c_{\Delta_\phi,0}$. There are no other infinities that this could cancel with until the infinite tower of heavier operators enters at $\oeps{2}$ in the bootstrap expressions; thus, we are \emph{forced} to set $C_\phi=\oeps{3}$ in order for the bootstrap equations to make any sense to $\oeps{2}$. Note that this suppresses the contribution of $\phi$ from all the bootstrap equations (\ref{eq:ll2})-(\ref{eq:l01}).

It is interesting to compare this with the $\phi^2$ case, as the $\phi$ $t$-channel contributions take an apparently similar form:
\begin{equation}\label{eq:divergenttermphi2}
c_{\Delta_\phi,0}q_{\Delta_\phi,l|0}^{(a,t)} \propto
	\begin{cases}
		\oeps{1} & d_0 = 6,\\
		C_\phi^{(0)}\Gamma\left(\frac{d_0-6}{4}\right)^{-2}\eps^{-1}(\infty) & d_0 > 6.
	\end{cases}
\end{equation}
However, let us emphasize that the reason $\phi$ must drop out of the bootstrap for $d_0>6$ is different than $\phi^2$: we saw that the contribution of the $\phi^2$ operator is automatically suppressed to $\oeps{2}$ for $d_0>4$; here, the contribution of $\phi$ in $d_0>6$ is actually \emph{infinite}, forcing us to set the OPE coefficient $C_\phi=\oeps{3}$. This could not have happened for the $\phi^2$ operator, as the $\phi\phi\phi^2$ OPE is already non-zero in the free theory.

\subsection{Bootstrap equations to \texorpdfstring{$\oeps{2}$}{O(e2)}}
Above, we have discussed the $t$-channel contributions from $\phi$ and $\phi^2$ and found that they all vanish. Now, we can use this to solve the bootstrap equations, which at this point consist solely of operators in the $s$-channel either cancelling off $t$-channel identity operator contributions (in the simple-pole constraints) or being set directly to $\oeps{2}$ (in the double-pole case).

Let us consider the $l=2$ bootstrap equations (\ref{eq:l22})-(\ref{eq:l21}) first. The lowest order term of (\ref{eq:l22}) is at $\oeps{1}$, while (\ref{eq:l21}) has terms at $\oeps{0}$ and $\oeps{1}$. Demanding that (\ref{eq:l22}) and (\ref{eq:l21}) hold up to $\oeps{1}$ thus gives us three equations to solve. The solution is
\begin{equation}\label{eq:solnl2}
	\delta_\phi^{(1)} = 0,\qquad C_2^{(0)} = C_{2,\text{free}}(0), \qquad C_2^{(1)} = C_{2,\text{free}}'(0),
\end{equation}
where $C_{2,\text{free}}(\eps)$ is the square of the (appropriately normalized) OPE coefficient in the free theory
\be \label{eq:C2free} C_{2,\text{free}}(\eps) = \frac{d_\eps}{16}\frac{(d_\eps-2)^2}{d_\eps - 1}, \qquad C'_{2,\text{free}}(0)= \left.\frac{\partial}{\partial\eps}C_{2,\text{free}}(\eps)\right|_{\eps=0}.\ee These results match those found in \cite{Gopakumar:2016cpb} for $d_\eps = 4-\eps$.

The $l\neq0,2$ bootstrap equations (\ref{eq:ll2})-(\ref{eq:ll1}) for $J_l$ proceed entirely analogously to the $l=2$ case. The results of the bootstrap give:
\be \delta_l^{(1)} = 0, \qquad C_l^{(0)} = C_{l,\text{free}}(0), \qquad C_l^{(1)} = C_{l,\text{free}}'(0),\ee
where:\footnote{Of course, (\ref{eq:Clfree}) agrees with (\ref{eq:C2free}) for $l=2$.}
\be \label{eq:Clfree} C_{l,\text{free}}(\eps) =  (-1)^l\frac{2}{l!} \frac{\Gamma^2\left(l + \frac{d_\eps}{2}- 1\right)\Gamma\left(l+d_\eps-3\right)}{\Gamma^2\left(\frac{d_\eps}{2}-1\right)\Gamma\left(d_\eps+2l-3\right)}.\ee

Next, we can consider the $l=0$ bootstrap equations (\ref{eq:l02})-(\ref{eq:l01}). Equation (\ref{eq:l01}) has terms at $\oeps{0}$ and $\oeps{1}$, which can be set to zero to obtain:
\begin{equation}\label{eq:solnl0}
	C_0^{(0)} = 2, \qquad C_0^{(1)} = 2 \delta_0^{(1)}\left( 2 H_{d_0/2-3} - H_{d_0-4} + \frac{3 d_0-8}{d_0^2-7 d_0+12}\right),
\end{equation}
where $H_n$ is the $n$-th harmonic number. This again matches \cite{Gopakumar:2016cpb} when $d_\eps = 4-\eps$ (for which $C_0^{(1)} = -2\delta_0^{(1)}$).

Finally, we turn to equation (\ref{eq:l02}), where the crucial distinction between $d_0=4$ and $d_0>4$ for the $\phi^2$ contributions plays a critical role. Using (\ref{eq:divergentterm}), we find that the $\oeps{1}$ term of (\ref{eq:l02}) gives us:
\begin{equation}\label{eq:delta01}
	\delta_0^{(1)} =
	\begin{cases}
		\frac13 & d_0 = 4\\
		0 & d_0 > 4
	\end{cases}
\end{equation} 
The $d_0=4$ value again matches that of~\cite{Gopakumar:2016cpb}. Because $c_{\Delta_0,0}q_{\Delta_0,0|0}^{(2,t)}$ is suppressed by an extra order of $\eps$ in $d_0>4$, we see that $\delta_0^{(1)}$ is forced to vanish in this case, giving the free theory result.

\subsection{Summary and Extensions}
To summarize, we have found that the quantities in question satisfy (for $d_0>6$):
\begin{align}
 \label{eq:delphisum} \Delta_\phi &= \Delta_{\phi,\text{free}} + \oeps{2} ,& \Delta_{\{0,2,l\}}& = \Delta_{\{0,2,l\},\text{free}} + \oeps{2},\\
\label{eq:Cphisum} C_\phi& = 0 + \oeps{3}, & C_{\{0,2,l\}} &= C_{\{0,2,l\},\text{free}} + \oeps{2}.
\end{align}
Therefore we conclude that any CFT in $d_0>6$ with a conserved stress tensor and a single fundamental scalar must be perturbatively free up to $\oeps{2}$. Furthermore, for $\mathbb{Z}_2$-symmetric theories the range of validity for these results extends to $d_0>4$, since the $\phi^2$ operator drops out already at that dimension. We stress that these results only depend on the value of $d_0$, and in particular not on the sign of $\eps$. These results can also be generalized to higher orders in $\eps$, as well as to include multiple scalars. 

\subsubsection*{Higher Orders in \texorpdfstring{$\eps$}{eps}}
While our results are certainly sugggestive, they are at relatively low orders in $\eps$. To what extent can we push this to deeper orders? We have found two regimes in which we can go further:
\begin{align}
\label{def:extraassumptions}
\begin{split}
	\bullet ~~d_\eps &= 4n + 2 + \oeps{1}, \text{ or}\\
	\bullet ~~d_\eps &= 2n + 2 + \oeps{1} \text{ and the CFT is $\mathbb{Z}_2$-symmetric},
\end{split}
\end{align}
for $n>1, n \in \mathbb{Z}^+$. As described in appendix \ref{sec:app:simpl} (see especially \ref{sec:app:coeffc}), when either of these is true, the simplifications \textbf{S1}-\textbf{S3} actually hold to $\oeps{4}$ (note this also requires $\delta_l^{(1)}=0$, which is guaranteed by (\ref{eq:delphisum})). This means that the right hand side of (\ref{eq:ll2})-(\ref{eq:l01}) is $\oeps{4}$ and we can then perform the bootstrap analysis above at orders $\oeps{2}$ and $\oeps{3}$ as well. The analysis of the equations at these orders proceeds entirely analogously as above, and results in determining the quantities in question to higher order in $\eps$:\footnote{Note that we expect the polynomial ambiguities in the Mellin space Witten diagrams in (\ref{eq:Ms}) and (\ref{eq:Mt}) to start contributing at $\oeps{4}$ \cite{Dey:2017fab}, which would complicate the analysis beyond this order in $\eps$ even further.}
\begin{align}
\label{eq:delphisumA3} \Delta_\phi &= \Delta_{\phi,\text{free}} + \oeps{4} ,& \Delta_{\{0,2,l\}}& = \Delta_{\{0,2,l\},\text{free}} + \oeps{4},\\
 \label{eq:CphisumA3} C_\phi& = 0 + \oeps{5}, & C_{\{0,2,l\}} &= C_{\{0,2,l\},\text{free}} + \oeps{4}.
\end{align}

\subsubsection*{Bootstrap for Multiple Scalars}\label{sec:multiplescalars}
Our analysis was done assuming only one fundamental scalar $\phi$ that saturates the unitary bound when the theory is free at $\eps=0$. However, it is fairly straightforward to see that we can relax this assumption to include an arbitrary number $N$ of fundamental scalars $\phi^{(i)}$, with dimensions:
\be \Delta_{\phi^{(i)}} = \frac{d_\eps-2}{2} + \delta_{\phi^{(i)}}^{(1)} \eps + \delta_{\phi^{(i)}}^{(2)} \eps^2 +\oeps{3}.\ee
We can perform $N$ versions of the four identical scalar bootstrap that we have performed above in the case of one scalar. The simplifications \textbf{S1}-\textbf{S3} still hold with arbitrary index structures on the heavy operators $\mathcal{O}_{k,l,m}$. The rest of the bootstrap proceeds analogously - for instance, the $l=2$ bootstrap equations will give e.g. $\delta_{\phi^{(i)}}^{(1)}=0$ for every $i$. The only possible extra non-triviality that one might worry about is the existence of a conserved spin-1 current (such as in the $O(N)$ model). However, since our results force the scalar theory into a free theory, the relevant spin 1 current is automatically conserved, anyway, without needing any further input.\footnote{Note that the spin-2 conserved current, the stress-energy current, is not automatically conserved. We need to demand its conservation in order to find the condition $\delta_{\phi^{(i)}}^{(1)}=0$.} Our analysis is thus insensitive to any demand of global symmetries the theory could have, as the resulting theory is always free.

We thus conclude that our analysis can be extended to the case of $N$ fundamental scalars $\phi^{(i)}$ with arbitrary indices placed on the relevant operators. It is surprising that with very little input or assumptions, we are able to see clear evidence that no perturbative interacting CFTs with \emph{any number of} fundamental scalars exist in $d_0>6$, without needing any notion at all of the Lagrangian. Perhaps by considering more of the infinite class of spurious poles in the Mellin bootstrap one could further relax our assumptions to fully explore the world of CFTs in $d_0>6$.

\section*{Acknowledgments}
We thank Marco Baggio, Anthony Charles, Ben Safdi, and Miguel Paulos for useful discussions, and especially Henriette Elvang for many helpful suggestions and guidance during this project. This work was supported by the U.S. Department of Energy under grant DE-FG02-95ER40899 and JG was supported in part by a Van Loo Postdoctoral Fellowship.

\appendix

\section{Definitions}\label{sec:app:def}
In this appendix, we collect a number of necessary and useful definitions and expressions we need; for more details and the derivations of these expressions, we refer the reader to \cite{Gopakumar:2016cpb} where they originally appeared. We will often use the notation:
\be h = \frac{d}{2},\ee
and the definition of the Pochhammer symbol:
\be (a)_b = \frac{\Gamma(a+b)}{\Gamma(a)}.\ee
The coefficient $c_{\Delta, l}$ is given by:
\be \label{eq:app:cNC} c_{\Delta, l} = C_{\Delta, l} \mathcal{N}_{\Delta, l},\ee
where $C_{\Delta,l}$ is the square of the OPE coefficient $C_{\phi\phi\mathcal{O}_{\Delta,l}}$, and $\mathcal{N}_{\Delta,l}$ is the normalization coefficient:
\be \left(\mathcal{N}_{\Delta,l}\right)^{-1} = \frac{\Gamma(\Delta-1)\Gamma^4\left(\frac{l+\Delta}{2}\right)}{(-2)^l(l+\Delta-1)\Gamma(1-h+\Delta)\Gamma^2(l+\Delta-1)} \Gamma^2\left(\frac{l-\Delta+2\Delta_\phi}{2}\right)\Gamma^2\left( \frac{\Delta+2\Delta_\phi-2h+l}{2}\right).\ee
The function $q^{(s)}_{\Delta,l}(s)$ is given by:
\be \label{eq:qsdef} q^{(s)}_{\Delta,l}(s) = -\frac{ 4^{1-l} \Gamma^2(\Delta_\phi + s + l-h)}{(l+2s-\Delta)(l+2s+\Delta-2h)\Gamma(2s+l-h)},\ee
from which the coefficients $q^{(2,s)}_{\Delta, l}$ and $q^{(1,s)}_{\Delta, l}$ can be extracted using:
\be \label{eq:appqscoef} q^{(2,s)}_{\Delta, l} =  q^{(s)}_{\Delta,l}(s=\Delta_\phi), \qquad q^{(1,s)}_{\Delta, l} = \frac{\partial}{\partial s}\left(q^{(s)}_{\Delta,l}(s)\right)_{s=\Delta_\phi}. \ee
These are given by:
\begin{align} q^{(2,s)}_{\Delta, l} &= -\frac{4^{1-l} \Gamma(2\Delta_\phi+l-h)}{(l-\Delta+2\Delta_\phi)(l+\Delta+2\Delta_\phi-2h)},\\
 q^{(1,s)}_{\Delta, l} &= \frac{ 4^{2-l} \Gamma(2\Delta_\phi+l-h+1)}{(l-\Delta+2\Delta_\phi)^2(l+\Delta+2\Delta_\phi-2h)^2}.
\end{align}
The function $q^{(t)}_{\Delta,l|l'}(s)$ is given by:
\begin{align}
 \nonumber q^{(t)}_{\Delta,l|l'}(s) &= \kappa_l(s)^{-1}\sum_{q=0}^l \widetilde{\sum}_{(l',k,m,\alpha,\beta)} \int d\nu\, \mu_{\Delta, l'}(\nu) \Gamma^2(\lambda_1)\Gamma^2(\overline{\lambda}_1) (\Delta_\phi-s)_{m-\alpha} (\Delta_\phi-s)_{l'-2k-m-\beta}\\
\nonumber & \times \frac{1}{\prod_{i=1}^4 \Gamma(l_i)} \frac{2^l (s)_l^2}{(2s+l-1)_l} \frac{ (-1)^q l! (2s+l-1)_q}{(s)_q^2 q!(l-q)!}\\
\nonumber & \times \frac{\Gamma(k+q+s+\alpha+\lambda_2-\Delta_\phi)\Gamma(k+q+s+\alpha+\overline{\lambda}_2-\Delta_\phi)}{\Gamma(q+2s+2k+\alpha+\beta+\lambda_2+\overline{\lambda}_2 -2\Delta_\phi)}\\
\nonumber & \times \Gamma(k+s+\beta-\Delta_\phi+\lambda_2)\Gamma(k+s+\beta-\Delta_\phi+\overline{\lambda}_2)\\
\label{eq:qtfull} & \times \phantom{h}_3F_2\left[\begin{array}{c} -q, -\alpha, 1-2k-q-2s-\alpha-\beta+2\Delta_\phi-\lambda_2-\overline{\lambda}_2\\ 1-k-q-s-\alpha-\lambda_2+\Delta_\phi, 1-k-q-s-\alpha-\overline{\lambda}_2+\Delta_\phi \end{array}; 1\right].
\end{align}
Here, we used the sub-expressions:
\begin{align}
 \mu_{\Delta,l'}(\nu) &= \frac{\Gamma^2(\Delta_\phi-\lambda_2)\Gamma^2(\Delta_\phi-\overline{\lambda}_2)}{2\pi i ((\Delta-h)^2-\nu^2)\Gamma(\nu)\Gamma(-\nu)(h+\nu-1)_{l'}(h-\nu-1)_{l'}},\\
 \kappa_l(s) &= \frac{(-1)^l 4^l l!}{(2s+l-1)_l^2} \frac{\Gamma^4(l+s)}{(2s+2l-1)\Gamma(2s+l-1)},
\end{align}
and the definition of the complicated summation:
 \begin{align}\nonumber \widetilde{\sum}_{(l',k,m,\alpha,\beta)} &= \frac{l'!}{2^{l'}(h-1)_{l'}} \sum_{k=0}^{\left[\frac{l'}{2}\right]} \sum_{m=0}^{l'-2k}\sum_{\alpha=0}^m\sum_{\beta=0}^{l'-2k-m} \frac{ (-1)^{l'-k-\alpha-\beta}\Gamma(l'-k+h-1)}{\Gamma(h-1)k!(l'-2k)!}\\
 & \times \left(\begin{array}{c}l'-2k\\ m\end{array}\right) \left(\begin{array}{c}m\\ \alpha\end{array}\right) \left(\begin{array}{c}l'-2k-m\\ \beta\end{array}\right),
 \end{align}
 and finally the shorthand notations:
\begin{align}
 \lambda_1 &= \frac{h+\nu+l'}{2}, & \overline{\lambda}_1 &= \frac{h-\nu+l'}{2}, & \lambda_2 &= \frac{h+\nu-l'}{2}, & \overline{\lambda}_2 &= \frac{h-\nu-l'}{2},
 \end{align}
 \begin{align}l_1 &= \lambda_2+l-k-m+\alpha-\beta, & l_2 &= \lambda_2+k+m-\alpha+\beta, &l_3 &= \overline{\lambda}_2+k+m, & l_4 &= \overline{\lambda}_2+l-k-m.
\end{align}
The expression (\ref{eq:qtfull}) simplifies considerably for $l'=0$ (with $\lambda\equiv\lambda_1=\lambda_2$ and $ \overline{\lambda}\equiv\overline{\lambda}_1=\overline{\lambda}_2$):
\begin{align}
 \nonumber q^{(t)}_{\Delta,l|l'=0}(s) &= \int d\nu\, \mu_{\Delta,0}(\nu) \frac{ 2^l ((s)_l)^2}{\kappa_l(s)(2s+l-1)_l}\Gamma(s - \Delta_\phi+\lambda)\Gamma(s-\Delta_\phi+\overline{\lambda})\\
 & \times \sum_{q=0}^{l} \frac{(-1)^q l! (s+l-1)_q}{((s)_q)^2 q!(l-q)!} \frac{ \Gamma(q+s+\lambda-\Delta_\phi)\Gamma(q+s+\overline{\lambda}-\Delta_\phi)}{\Gamma(q+2s+\lambda+\overline{\lambda}-2\Delta_\phi)}
\end{align}
The coefficients $q^{(2,t)}_{\Delta, l|l'}$ and $q^{(1,t)}_{\Delta, l|l'}$ can be extracted from $q^{(t)}_{\Delta,l|l'}(s)$ in a similar way as in (\ref{eq:appqscoef}).
For completeness, we also give the expression for the continuous Hahn polynomials $Q_l^\Delta(t)$, which we give in a convenient form:
\be Q_l^{2\Delta_\phi+l}(t) = \frac{2^l (\Delta_\phi)^2_l}{(2\Delta_\phi+l-1)_l} \phantom{h}_3F_2\left[ \begin{array}{c} -l, 2\Delta_\phi+l-1, \Delta_\phi+t\\ \Delta_\phi, \Delta_\phi \end{array}; 1\right].\ee
These satisfy the orthonormality condition:
\be \frac{1}{2\pi i} \int_{-i\infty}^{+i\infty} dt\, \Gamma^2(\Delta_\phi+t)\Gamma^2(-t)Q_l^{2\Delta_\phi+l}(t) Q_{l'}^{2\Delta_\phi+l'}(t) = \kappa_l(\Delta_\phi)\delta_{l,l'}.\ee

\section{Simplifications}\label{sec:app:simpl}

Here we give more information about the simplifications \textbf{S1} - \textbf{S3} (the simplification \textbf{S0} relies only on the fact that we are working with identical scalars, therefore the derivation given in \cite{Gopakumar:2016cpb} follows through). These simplifications deal with the contributions of generic operators $\mathcal{O}_{k,2m,l}$ that contains $k$ $\phi$'s, $l$ uncontracted derivatives, and $2m$ contracted derivatives. Such an operator will have spin $l$ and dimension:
\be \label{eq:app:heavyDelta} \Delta_{\mathcal{O}_{k,2m,l}} = k\frac{d_{\eps}-2}{2} + l + 2m + \delta^{(1)}_{k,2m,l}\eps + \delta_{k,2m,l}^{(2)}\eps^2 + \oeps{3}.\ee
For a given $k,m,l$, one possible such operator is (schematically) $\phi^{k-1}\partial_{a_1}\cdots\partial_{a_l}(\partial^2)^m\phi$, but there could possibly be other operators (by distributing the derivatives differently among the $\phi$'s).\footnote{Note that, in \cite{Gopakumar:2016cpb}, only operators with $k=2$ were possible; this is because they explicitly consider the $\phi^4$ theory where $\phi^3$ is a descendant operator. Note that we could possibly be ``overcounting'' the number of operators when considering generic $\mathcal{O}_{k,2m,l}$ (i.e. also considering some descendants), but this is irrelevant since we prove we can ignore their contribution to the bootstrap equations anyway.} However, \emph{any} such operators $\mathcal{O}_{k,2m,l}$ and $\mathcal{O}_{k,2m,l}'$ will only differ by the value of the coefficients $\delta_{k,2m,l}^{(i)}$.

\subsection{The coefficients \texorpdfstring{$c_{\Delta_{\mathcal{O}},l}$}{c}}\label{sec:app:coeffc}
First of all, let us consider the coefficients $c_{\Delta_{\mathcal{O}},l}$ for the heavy operators $\mathcal{O}_{k,2m,l}$ with dimension (\ref{eq:app:heavyDelta}). These coefficients are given by (\ref{eq:app:cNC}); we discuss the factor $C_{\phi\phi\mathcal{O}_{k,2m,l}}$ and $\mathcal{N}_{\Delta,l}$ separately.

In the free theory, the three point function of $\phi\phi \mathcal{O}_{k,2m,l}$ for $k>2$ vanishes (because e.g. the three point function $\phi\phi\phi^k$ vanishes for $k>2$); this implies that:
\be \label{eq:app:Ccoeffeps2} C_{\phi\phi\mathcal{O}_{k,2m,l}}= \oeps{2}, \ee
(or a higher power) as the $C$ coefficient is the \emph{square} of the $\phi\phi \mathcal{O}_{k,2m,l}$ OPE coefficient (which is reasonable to assume goes as $\oeps{1}$). For $k=2$, we can consider only (primary) operators where the derivatives hit only one of the $\phi$'s, i.e. of the schematic form $\phi (\partial^2)^m \partial_{i_1}\cdots \partial_{i_l}\phi$. For $m>0$, these operators identically vanish in the free theory so that we again have (\ref{eq:app:Ccoeffeps2}). Only the operators with $k=2, m=0$ (i.e. the $J_l$ operators) may have:
\be \label{eq:app:Ccoeffeps0} C_{\phi\phi\mathcal{O}_{2,0,l}} = \oeps{0}.\ee

Next, we note that the normalization factor $\mathcal{N}_{\Delta,l}$ of every heavy operator with $k=2$ goes as:\footnote{We already set $\delta_{\phi}^{(1)}=0$ here for simplicity; this does not change the leading power of $\eps$ in $\mathcal{N}_{\Delta,l}$.}
\be \label{eq:norml} \mathcal{N}_{\Delta,l} = \left(\delta^{(1)}_{k=2,2m,l}\right)^2 g(h_0) \oeps{2} + \oeps{4},\ee
while when $k>2$, we have:
\be \mathcal{N}_{\Delta,l} =  \Gamma\left(-\frac12(k-2)(h_0-1)-m\right)^{-2}\tilde{g}(h_0) \oeps{0},\ee
where $g(h_0)$ and $\tilde{g}(h_0)$ are unimportant additional factors involving $h_0$. Note that the inverse gamma function has a zero when its argument is a negative integer; the only way to achieve this here is if $k=2n_1,h_0=n_2$ with $n_1,n_2$ integers, or if $h_0=2n+1$ with $n$ integer (since $k$ is always an integer). Remembering that $d=2h$, this is precisely the condition presented in (\ref{def:extraassumptions}). So, when (\ref{def:extraassumptions}) holds, we have:
\be \label{eq:normA3} \mathcal{N}_{\Delta,l} = \oeps{2}.\ee

Putting everything together, we can conclude that for any heavy operator $\mathcal{O}_{k,2m,l}$, we have:
\be \label{eq:app:ceps2} c_{\Delta_{\mathcal{O}},l} = \oeps{2}.\ee
When (\ref{def:extraassumptions}) holds and when $\delta^{(1)}_{2,0,l} = \delta^{(1)}_l = 0$, (\ref{eq:normA3}) and (\ref{eq:norml}) imply that we have:
\be \label{eq:app:ceps4} c_{\Delta_{\mathcal{O}},l} = \oeps{4}.\ee

\subsection{S1: \texorpdfstring{$s$}{s}-channel contributions for heavier operators}
Expanding (\ref{eq:qsdef}) around $\eps=0$ for heavier operators $\mathcal{O}_{k,2m,l}$ with $k>2$ and/or $m>0$, gives 
\begin{equation}
	q^{(s)}_{\Delta,l}(s) = \oeps{0}.
\end{equation}
Since (\ref{eq:app:ceps2}) holds for these operators, heavier operators in the $s$-channel begin contributing at $\oeps{2}$, i.e. we have:
\be c_{\Delta_{\mathcal{O}},l} q^{(a,s)}_{\Delta_\mathcal{O},l} = \oeps{2}. \ee
When (\ref{def:extraassumptions}) holds and $\delta^{(1)}_l = 0$, this becomes $c_{\Delta_{\mathcal{O}},l} q^{(a,s)}_{\Delta_\mathcal{O},l} = \oeps{4}$.

\subsection{S2: \texorpdfstring{$t$}{t}-channel contributions for \texorpdfstring{$l'>0$}{l'>0}}
For the simplification \textbf{S2}, we want to prove that operators $\mathcal{O}_{k,2m,l'}$ for $l'>0$ (with $k\geq 2$ and $m\geq 0$) satisfy:
\be \label{eq:S2maininapp} c_{\Delta_{\mathcal{O}}, l'} q_{\Delta_{\mathcal{O}},l|l'}^{(a,t)} = \oeps{2},\ee
for $l=0$ and $l=2$.\footnote{In checking these simplifications for $l\neq2$, it is computationally favorable to already set $\delta_\phi^{(1)}=0$ (as briefly mentioned in \cite{Gopakumar:2016cpb}). The only place where we need to keep $\delta_\phi^{(1)}$ explicitly non-zero is in checking the simplifications for $q^{(2,t)}$ when $l=2$.}

We first note that the integrand in the integral expression for $q_{\Delta_{\mathcal{O}},l|l'}^{(a,t)}$ (for $a=1,2$) has the following poles:
\begin{itemize}
 \item[I.] $\nu = \Delta_\mathcal{O}-h$
 \item[II.] $\nu = 2\Delta_\phi + l' - h + 2n_2$
 \item[III.] $\nu = h +l'+ 2n_3$ 
 \item[IV.] $\nu = h-1+n_4$; $\qquad$ $(n_4<l')$
\end{itemize}
In poles II and III, $n_2,n_3, $ can be any non-negative integer; these are infinite families of poles; for the poles IV, $n_4$ is a non-negative integer in the range $0,1,\cdots, l'-1$. The integral expression for $q_{\Delta_{\mathcal{O}},l|l'}^{(a,t)}$ will then have contributions from the residues at each of these poles, but the residues can be seen to combine to give vanishing contributions up to $\oeps{0}$ (for both $q^{(1,t)}$ and $q^{(2,t)}$ separately). The cancellations that occur for any $\mathcal{O}_{k,2m,l'}$ are:
\begin{align}
\label{eq:S2polecan1} Res_{II(n_2=n_3+1)} + Res_{III(n_3)} &= \oeps{0}, & n_3\neq m-1 ,\\
\label{eq:S2polecan2} Res_{IV(n_4)} &= \oeps{0}, & n_4=0,1,\cdots,l'-2 .
\end{align}
The rest of the poles cancel in slightly different ways for $m=0$ and $m>0$:
\begin{align}
\label{eq:S2polecan3} Res_I + Res_{II(n_2=0)} + Res_{IV(n_4 = l'-1)} &= \oeps{0} & (m=0),\\
\label{eq:S2polecan4} Res_I +  Res_{II(n_2=m)} + Res_{III(n_3=m-1)} &= \oeps{0} & (m>0),\\
\label{eq:S2polecan5} Res_{II(n_2=0)} + Res_{IV(n_4=l'-1)} &= \oeps{0} & (m>0).
\end{align}
Thus, we can conclude that $q_{\Delta_{\mathcal{O}},l|l'}^{(a,t)}= \oeps{0}$. Together with (\ref{eq:app:ceps2}), this proves that (\ref{eq:S2maininapp}) holds. When (\ref{def:extraassumptions}) holds and $\delta^{(1)}_l = 0$, (\ref{eq:app:ceps2}) implies that (\ref{eq:S2maininapp}) becomes suppressed to $\oeps{4}$.

\subsection{S3: \texorpdfstring{$t$}{t}-channel contributions for heavier scalars (\texorpdfstring{$l'=0$}{l'=0})}
For the simplification \textbf{S3}, we want to show that for operators $\mathcal{O}_{k,2m,l'}$ with $l'=0$ (i.e. scalar operators) that are heavy, i.e. $k>2$ and/or $m>0$, we have:
\be \label{eq:S3maininapp} c_{\Delta_{\mathcal{O}}, l'=0} q_{\Delta_{\mathcal{O}},l|l'=0}^{(a,t)} = \oeps{2}.
\ee
The analysis proceeds in a similar way to \textbf{S2}. The poles I, II, III (with $l'=0$) are present in the integrand for $q^{(a,t)}$, while the poles IV are now absent. The pole cancellations (\ref{eq:S2polecan1}) still holds. For $m>0$, the cancellations (\ref{eq:S2polecan4}) still hold. The final poles present satisfy:
\begin{align}
 \label{eq:S3polecan1} Res_I + Res_{II(n_2=0)} &= \oeps{0} & (m=0),\\
 \label{eq:S3polecan2} Res_{II(n_2=0)} &= \oeps{0} & (m>0).
 \end{align}
We again have (\ref{eq:app:ceps2}), so we can conclude that (\ref{eq:S3maininapp}) indeed holds. When (\ref{def:extraassumptions}) holds and $\delta^{(1)}_l = 0$, (\ref{eq:app:ceps2}) implies that (\ref{eq:S3maininapp}) becomes suppressed to $\oeps{4}$.

\subsection{Contributions from \texorpdfstring{$\phi$}{phi} and \texorpdfstring{$\phi^2$}{phi2}}
For the operators $\phi$ and $\phi^2$ in the $t$-channels, the cancellations (\ref{eq:S2polecan1}) again holds for $q_{\Delta_{\{\phi,\phi^2\}},l|0}^{(a,t)}$. The only other possible contributions can come from $Res_I$ and $Res_{II(n_2=0)}$. For $d_0$ larger than the special dimension $4$ (for $\phi^2$) or $6$ (for $\phi$), these are also suppressed as in (\ref{eq:S3polecan1}), so that all of the poles of $q_{\Delta_{\{\phi,\phi^2\}},l|0}^{(a,t)}$ go as $\oeps{0}$. However, there is a discontinuity for $q_{\Delta_{\{\phi,\phi^2\}},0|0}^{(2,t)}$ in this cancellation for $d_0=4$ (for $\phi^2$), as discussed in sec. \ref{sec:bootstrapphisq}. For $\phi$ there are in fact discontinuities at both $d_0=4$ (for $Res_{II(n_2=0)}$) and $d_0 = 6$ (for $Res_I$). However these $\phi$ discontinuities end up playing no significant role as the $\phi$ operator does not contribute for the $\mathbb{Z}_2$ case at $d_0=4$ and then, as discussed in sec. \ref{sec:bootstrapphi}, the normalization coefficient $c_{\Delta_\phi,0}=\oeps{1}$ in $d_0=6$.

\bibliographystyle{utphys}
\bibliography{MellinLargeD}

\end{document}